# Navigating Culture in Smart Port Cities:

# Cultural Sensitivity and Digital Engagement Among Sailing Tourists in the Mediterranean

Panagiota konstantinou
*Hellenic Army Academy,*
pkonstantinou@uniwa.gr

Georgios Stathakis
*Hellenic Army Academy,*
georgios.stathakis@ouc.ac.cy

## Abstract

This study examines the relationship between smart port city infrastructure, tourists/crew cultural sensitivity and digital engagement among international sailing tourists in the Mediterranean and particularly in Greece. It is based on an interdisciplinary literature synthesis and primary data from a survey conducted with a total of 203 respondents over three sailing seasons (2023–2025). This paper proposes a conceptual framework that positions cultural sensitivity as a result of the interaction between smart port destination technology, tourist awareness and their engagement with the local community. Among the findings, high levels of adoption of digital platforms for logistical purposes such as (navigation apps: 89.7%; weather apps: 86.7%), while culturally oriented digital tools remain underused (20.2%). A significant discrepancy is found between tourists' cultural sensitivity (91.1% rate cultural respect as very important) and their practical uncertainty in real cultural situations (62.6%). Thus highlighting an unmet need of potential visitors for real-time cultural guidance tools. Tourists from distant cultures found to be significantly higher among the entire sample of tourists. The evidence that tourists seek a culturally integrated smart port application is strong, particularly among tourists who experienced the highest levels of uncertainty (91.2%). The study contributes both conceptual and empirical evidence to the smart cities literature, with practical implications for port planners, tourism policy makers, and digital platform designers.

## 1. Introduction

The Mediterranean region is a cradle of civilization, cultural exchange and maritime heritage. More specifically, Greece is one of the most popular sailing destinations in the Mediterranean region, attracting a diverse flow of incoming tourists who navigate its coasts, anchor in its historic ports and interact with the local communities that inhabit them. Today, ports are increasingly investing in smart city technologies to manage tourist flows, improve their logistical infrastructure and efficiency. However, there is a risk of cultural sensitivity being overlooked from the holistic design of the touristic experience.

Smart cities are defined to the use of Internet of Things (IoT) technologies, data analytics, and digitally designed platforms to improve urban living and quality of life (Sanchez-Corcuera et al., 2019; Talebkhah et al., 2021). In coastal and port cities, these paradigms are increasingly being applied to tourism management, digital marina services, smart dock systems, real-time visitor information platforms, and digital mobile applications. The logistical benefits of such systems are well documented, but their potential to foster a culturally sensitive and community-aware tourism experience has received comparatively less scholarly attention.

This gap plays an important role in the context of sailing tourism. Unlike passengers on large cruise ships or tourists who choose to stay in land-based resorts, sailing tourists tend to visit smaller, less commercially developed ports, engaging more directly with local infrastructure, markets and residents. Their interactions are closer and less mediated by official tourism structures, making them more exposed to cultural friction between tourists and local communities when expectations or rules are not clearly communicated.

As sailing tourism continues to internationalize - driven by the expansion of yacht charter and the accessibility of Mediterranean itineraries to non-European tourists - the cultural dimension of these encounters is also increasing. However, tourists from culturally distant backgrounds may also have limited knowledge of Mediterranean customs, religious practices, and related environmental norms. Furthermore, when there is insufficient cultural guidance, even the most well-intentioned tourists may inadvertently offend or even completely disconnect from local community interaction.

This study argues that smart port city technologies have the potential to bridge this gap, by incorporating culturally sensitive information into digital platforms offered in ports and marinas. In this way, smart port cities could improve both the tourist experience and strengthen the relationship between visitors and host communities. The central objective of this study is to examine the relationship between the digital services of smart port cities and the cultural sensitivity and engagement of potential tourists.

There are three specific objectives that guide this research:

1. To develop a comprehensive conceptual framework linking smart port technology to tourists' cultural sensitivity and engagement with the local community.
2. To assess tourists' level of digital tool usage and their corresponding cultural awareness through a primary survey with a questionnaire collected in three different sailing seasons and for three consecutive years.
3. To identify potential gaps between tourists' self-reported cultural sensitivity and the practical uncertainty they may experience in potential cultural situations and finally to assess the demand for culturally embedded smart port applications.

The paper is organized as follows: Section 2 reviews the relevant literature. Section 3 presents the conceptual framework. Section 4 describes the methodology followed in the research. Section 5 presents the results. Section 6 discusses the findings. Section 7 outlines the policy implications. Section 8 highlights the limitations and future directions. Section 9 provides the conclusions.

## 2. Literature Review

## 2.1 Smart cities and port infrastructure

The concept of smart cities has been evolving continuously over the past two decades, shifting towards a technology-driven paradigm and a more holistic approach that integrates infrastructure, governance and citizen participation (Cowley et al., 2018; Datta, 2017). Smart cities leverage IoT technologies, big data analytics and purpose-built digital platforms to enhance governance and resource allocation (Sanchez-Corcuera et al., 2019; Talebkhah et al., 2021). In port environments, these capabilities relate to ship traffic management, berth allocation, environmental monitoring and tourist information services. However, cultural and social dimensions are largely neglected in the literature on smart port design, particularly in relation to the tourist experience.

## 2.2 Participatory sensing and digital tourism

Participatory sensing and real-time data collection through geo-located devices owned by active citizens — is increasingly emerging as a key mechanism for improving urban responsiveness and management (Burke et al., 2006; Konstantinou et al., 2021a). In tourism contexts, digital platforms have transformed visitor behavior, enabling real-time navigation, booking, and access to destination-related information. Recent studies suggest that tourists increasingly rely on digital tools for both trip planning and decision-making once they are there. However, the development of tools related to cultural guidance remains at an early stage, creating a gap in both practice and research.

## 2.3 Cultural sensitivity in tourism

When we refer to cultural sensitivity in tourism, we are talking about the awareness, respect and adaptation of tourist behavior in relation to prevailing local cultural norms, values and practices (Li, et al., 2020). This research identifies cultural distance and the difference between a tourist's culture and the culture of the  inhabitants country — as a factor that increases cultural uncertainty and potential friction between them (Mota-Rojas et al., 2021). Previous exposure and familiarity with the destination are equally important factors. Despite its importance, cultural sensitivity is rarely incorporated into the design of smart cities or digital tourism platforms, a gap that this study seeks to address.

## 2.4 Mediterranean sailing tourism

Sailing tourism in the Mediterranean, and specifically in Greece, holds a unique position in the broader cruise and maritime tourism sector. The country's fleet is mainly composed of smaller vessels, for flexible itineraries that also involve more direct community involvement. It differs significantly from mass cruise tourism in terms of its social and cultural footprint. The rich archaeological, gastronomic and cultural heritage of the Mediterranean makes it a special destination for examining the intersection between tourism and cultural sensitivity. However, despite the developed yacht charter market and the tourist internationalization of the sailing industry, the corresponding empirical research dealing with the cultural behavior of sailing tourists remains limited.

## 3. Conceptual Framework

This study proposes a conceptual framework that positions culturally sensitive sailing tourism as a product of three mutually reinforcing pillars. As illustrated in Figure 1, these are: (1) smart port technologies, which include IoT infrastructure, digital information platforms, and mobile phone applications; (2) cultural sensitivity of tourists, which includes both cultural awareness of the destination they are visiting and a willingness to engage with local norms and a corresponding demand for real-time guidance; and (3) local community engagement, trust-building, culturally appropriate communication, and relevant participatory governance mechanisms.

This integrated framework suggests that no single pillar is sufficient on its own. Smart port technology without cultural content offers only logistical efficiency. Cultural sensitivity without corresponding digital support leaves tourists without the guidance they need. And community engagement without technological infrastructure limits the expansion and reach of local cultural knowledge. Maximum effectiveness is achieved only when all three pillars above are equally aligned and mutually reinforcing, producing what this study calls culturally integrated smart port tourism.

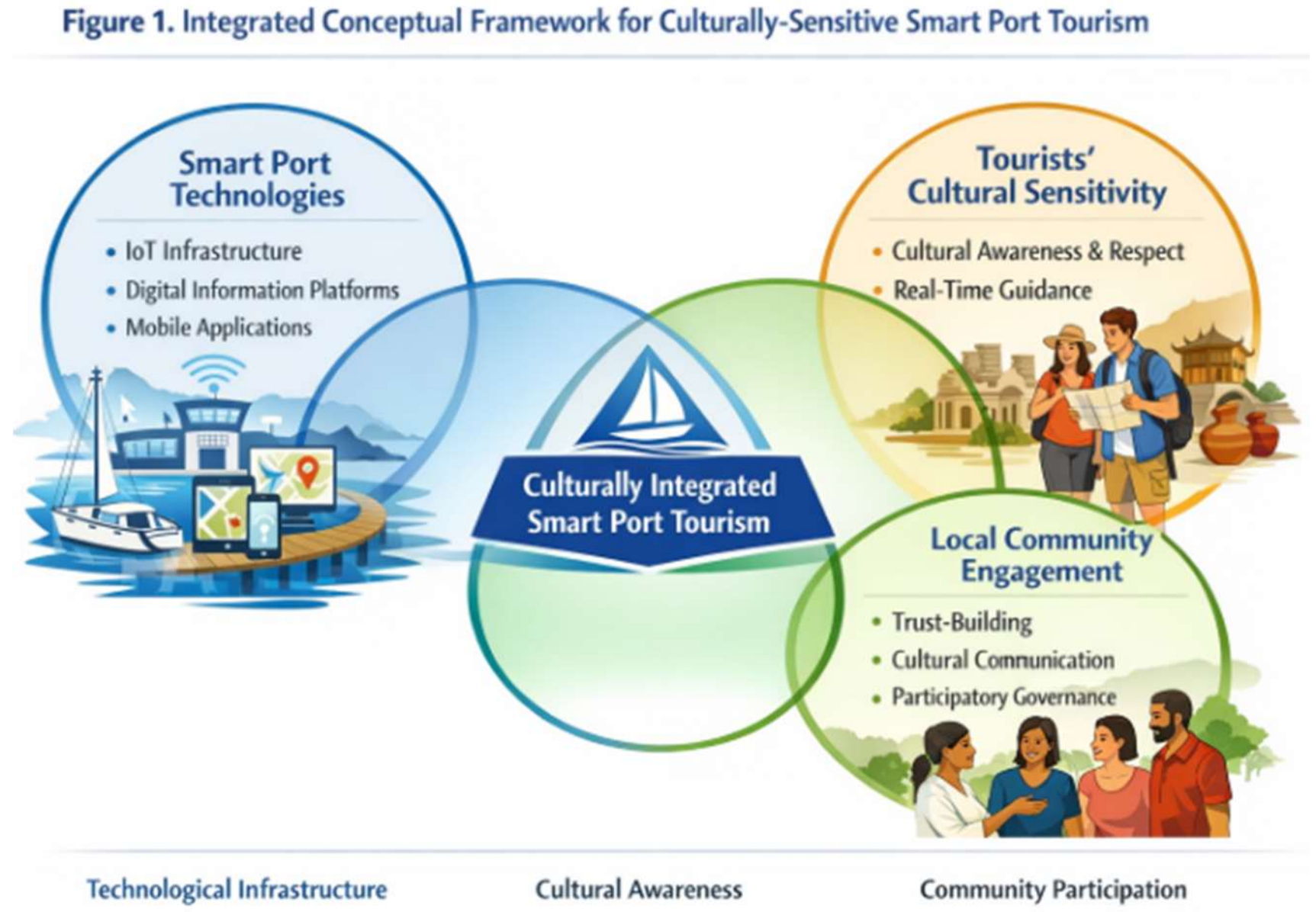

*Figure 1. Integrated conceptual framework for culturally-sensitive smart port tourism*

---

## 4. Methodology

This study adopts a quantitative research methodology. A structured questionnaire of 13 questions was distributed to international sailing tourists through a yacht charter company operating in Greek waters. The questionnaire had four distinct thematic sections: (i) the demographic profile of tourists and their respective travel experience, (ii) the use of digital tools and their relative satisfaction with smart port destinations, (iii) cultural awareness and/or uncertainty, and (iv) cultural engagement between visitors and the community and the corresponding demand for smart port applications. The questions were multiple choice or five-point Likert scale.

Data collection was carried out in three sailing seasons: between the years 2023, 2024 and 2025, covering departures throughout the tourist season, i.e. in spring, summer and early autumn, in order to minimize the possibility of seasonal bias. The questionnaires were administered in hard copy to the clients after receipt of the charter boats and before their departure, resulting in a high response rate estimated at over 80%. A total of 203 completed questionnaires were retained for analysis after excluding incomplete submissions.

Descriptive statistics including frequency distributions, percentage responses and mean Likert scores were calculated for all questionnaires. Cross-tabulation analysis of all tables was also conducted to examine variation in satisfaction and cultural uncertainty by age group, region of origin and frequency of visit. Analyses were conducted using standard statistical procedures used in comparable survey-based tourism research (Li, et al., 2020).

The study was conducted in accordance with the ethical principles of voluntary participation and anonymity. No personally identifiable information was collected and participants were informed of the purpose of the study before completing the questionnaire.

## 5. Results

### 5.1 Sample distribution

The survey data set consists of 203 complete responses collected during the tourist seasons (April - November) 2023 – 2025. As shown in Table 1, the distribution of respondents is relatively balanced over the three years, with 30.5% collected in 2023, 35.0% in 2024 and 34.5% in 2025. This distribution supports the temporal consistency of the data set by reducing the possibility of bias by year.

**Table 1.** *Survey distribution by year*

| Year | Respondents (N) | Percentage (%) |
|---|---|---|
| 2023 *June to November* | 62 | 30.5 |
| 2024 *April to November* | 71 | 35.0 |
| 2025 *April to November* | 70 | 34.5 |
| **Total** | **203** | **100.0** |

*Note. N = 203. Data collected across three sailing seasons April to November.*

### 5.2 Digital application usage

Table 2 presents the distribution of digital application usage among the sample (sailing tourists). These findings reveal a high level of adoption of sailing but general-purpose applications, with 89.7% of respondents using navigation applications and 86.7% using weather applications. Booking applications are also widely used (73.4%). In contrast, only 20.2% of respondents reported using applications specifically designed for cultural guidance. This result highlights a gap between the widespread use of digital tools by the sailing community mainly for navigation and weather information purposes and not for culturally oriented digital services, thus underlining the unmet demand for culturally integrated smart port applications.

**Table 2.** *Use of digital applications by sailing tourists*

| Type of application | Users (N) | Percentage (%) |
|---|---|---|
| Navigation apps | 182 | 89.7 |
| Weather apps | 176 | 86.7 |
| Booking apps | 149 | 73.4 |
| Cultural guidance apps | 41 | 20.2 |

*Note. Multiple responses permitted. Percentages reflect proportion of total respondents (N = 203).*

### 5.3 Satisfaction with smart port digital services

Table 3 presents the average satisfaction levels with smart port digital services on a five-point Likert scale. Overall satisfaction is moderate, with a mean score of 3.18. Individual service components, such as Wi-Fi availability (mean = 3.4), online booking systems (mean = 3.2), and digital information displays (mean = 3.1), indicate acceptable but not high levels of user satisfaction.

**Table 3.** *Satisfaction with smart port digital services (1–5 scale)*

| Service category | Mean score (1–5) |
|---|---|
| Wi-Fi availability | 3.4 |
| Online booking systems | 3.2 |
| Digital information displays | 3.1 |
| Mobile services / apps | 3.0 |
| **Overall average** | **3.18** |

*Note. 1 = very dissatisfied; 5 = very satisfied.*

Table 4 provides further age-group analysis and reveals a decreasing trend in satisfaction among older users. Respondents aged 18–35 report the highest satisfaction (mean = 3.4), while those aged 56 and over report lower levels (mean = 2.9). This suggests that the level of usability, ease of accessibility and digital familiarity of users may influence their perception of smart port services.

**Table 4.** *Satisfaction with smart port digital services by age group*

| Age group | Mean satisfaction (1–5) |
|---|---|
| 18–35 | 3.4 |
| 36–55 | 3.2 |
| 56 and above | 2.9 |

*Note. 1 = very dissatisfied; 5 = very satisfied.*

### 5.4 Cultural awareness and practical uncertainty

Table 5 examines the relationship between self-reported cultural sensitivity and actual experience of respondents. The vast majority of respondents (91.1%) state that cultural respect is very important to them, indicating a high level of stated cultural awareness. However, only 54.2% report feeling confident about understanding local customs, while 62.6% acknowledge that they are uncertain when faced with real-life situations. This reveals a notable discrepancy between intention and practical readiness and is a key empirical finding of this study.

**Table 5.** *Cultural attitudes and practical experience*

| Statement | Agree (%) |
|---|---|
| Cultural respect is very important to me | 91.1 |
| I felt confident about local customs | 54.2 |
| I felt uncertain in some situations | 62.6 |

*Note. Percentage of respondents agreeing or strongly agreeing with each statement (N = 203).*

### 5.5 Influence of nationality and prior experience

Table 6 presents the cultural uncertainty of respondents by region of origin. Tourists from Mediterranean countries report the lowest uncertainty (mean = 2.4), while those from Northern Europe (mean = 3.2), North America (mean = 3.5) and Australia (mean = 3.6) report significantly higher levels. Table 7 shows that first-time visitors to the destination show significantly higher uncertainty (mean = 3.7) compared to repeat visitors (mean = 2.6). The above findings suggest that both cultural distance and familiarity with the destination are factors that influence cultural readiness.

**Table 6.** *Cultural uncertainty by region of origin*

| Region of origin | Mean uncertainty score (1–5) |
|---|---|
| Mediterranean countries | 2.4 |
| Northern Europe | 3.2 |
| USA / Canada | 3.5 |
| Australia | 3.6 |

*Note. 1 = no uncertainty; 5 = very high uncertainty.*

**Table 7.** *Cultural uncertainty by visit frequency*

| Visitor type | Mean uncertainty score (1–5) |
|---|---|
| First-time visitors | 3.7 |

| Repeat visitors | 2.6 |
|---|---|

*Note. 1 = no uncertainty; 5 = very high uncertainty.*

## 5.6 Patterns of cultural engagement

Table 8 presents the types of cultural activities undertaken by the respondents. The results show that tourists are mainly involved in gastronomic culture (88.2%) and visits to historical-archaeological sites (81.3%). Participation in organized local cultural events is moderate (54.7%). In contrast, direct interaction with local communities is significantly lower (29.6%), indicating that tourists' cultural engagement is largely passive rather than participatory. This pattern suggests that visitors tend to consume culture as an experience rather than actively participate in local social environments.

**Table 8.** *Types of cultural engagement*

| Cultural activity | Participation (%) |
|---|---|
| Food and gastronomy | 88.2 |
| Historical sites | 81.3 |
| Cultural events | 54.7 |
| Local community interaction | 29.6 |

*Note. Multiple responses permitted. Percentages reflect proportion of total respondents (N = 203).*

## 5.7 Demand for culturally-oriented smart port applications

Table 9 presents the interest of tourists in a culturally targeted smart port application. The majority of respondents express their willingness to use such a tool, with 64.5% indicating definite interest and an additional 23.6% expressing possible interest, resulting in a combined positive response rate of 88.1%. Table 10 shows that the demand for this type of digital application concerns tourists who reported higher levels of cultural uncertainty (91.2%), compared to 68.3% among those with lower uncertainty. This alignment between the perceived need of visitors and the corresponding technological acceptance strongly supports the need for the development of culturally integrated digital services.

**Table 9.** *Interest in a culturally-aware smart port application*

| Response | Percentage (%) |
|---|---|
| Yes, definitely | 64.5 |
| Yes, probably | 23.6 |
| Not sure | 8.4 |
| No | 3.5 |

*Note. N = 203.*

**Table 10.** *App demand by level of cultural uncertainty*

| Group | Willing to use app (%) |
|---|---|
| Tourists with high cultural uncertainty | 91.2 |
| Tourists with low cultural uncertainty | 68.3 |

*Note. High uncertainty defined as mean score ≥ 3.5 on the cultural uncertainty scale.*

## 6. Discussion

The findings of the study show a series of interrelated trends at the heart of sailing tourism. Overall, they reveal a system that is digitally capable but culturally underprepared. In essence, it is a system in which the infrastructure for digital engagement has overtaken the development of tools for cultural guidance and community connection.

The most striking finding concerns the use of digital applications. The high adoption of navigation and weather forecasting applications (89.7% and 86.7% respectively) confirms that sailing tourists are highly digitally engaged. However, only one in five (20.2%) uses any form of culturally oriented digital guidance. This asymmetry does not only arise from tourists' preferences but also reflects a gap in the provision of culturally integrated digital services in Greek ports. As has been argued in relation to participatory exploration more broadly, the availability and design of digital platforms directly shape the behaviours they enable (Konstantinou et al., 2021). Therefore, the absence of culturally focused digital applications constitutes a structural barrier to cultural engagement.

A second key finding is the discrepancy between cultural sensitivity and cultural uncertainty practice. The fact that 91.1% of respondents consider cultural respect to be very important, while 62.6% simultaneously acknowledge having felt cultural uncertainty, suggests that visitors' good intentions alone are not sufficient for effective intercultural interaction. This finding aligns with a growing body of tourism research that identifies the gap between dimensions of cultural sensitivity, such as those based on attitude and behavior (Li, J., et al., 2020; Mota-Rojas et al., 2021). It further highlights the need for real-time, context-based cultural guidance that a culturally integrated smart tourism application could provide.

Data on nationalities and experiences support this argument. Significantly higher uncertainty scores among visitors from non-Mediterranean backgrounds (North America: 3.5; Australia: 3.6) and among first-time visitors (3.7) confirm that cultural distance and unfamiliarity are the two main factors that enhance cultural uncertainty. The above patterns are consistent with the cross-cultural literature on tourism and suggest that culturally targeted digital tools will have the greatest impact.

The findings on cultural engagement reveal a pattern of passive cultural consumption. The high participation rates in gastronomy (88.2%) and visits to cultural heritage sites (81.3%) are in direct contrast to the low rate of interaction with the local community (29.6%). This pattern of superficial engagement may, however, reflect both a lack of trust in different cultural norms and a lack of digital infrastructure to facilitate deeper connection with the community. Smart city applications that provide cultural information and actively facilitate community interaction could positively contribute to shaping this balance towards a more active and reciprocal cultural exchange.

Finally, the strong demand for smart applications targeting cultural awareness (88.1% combined positive responses) provides a clear empirical validation for the conceptual framework proposed in this study. Importantly, demand is highest among those who need it most, suggesting that such tools would not only be used, but would address a real need. This alignment between demand and need provides a strong foundation for policy investment.

## 7. Policy and Practical Implications

### 7.1 For smart port city planners

The findings of this study provide clear guidance on the expansion of digital applications to cultural guidance issues. Current investments in smart port applications in Greece are based on logistical efficiency, berth management, ship tracking and connectivity infrastructure. While these services are valuable, they only cover a part of the overall tourist experience. Thus, port authorities should invest in culturally integrated digital platforms that provide targeted information on local customs, religious practices, events, etiquette rules easily and directly accessible through smart tourist port applications.

### 7.2 For tourism policymakers

Stakeholders should integrate cultural sensitivity tools into their tourism strategies. This could include developing multilingual cultural guidance applications, ensuring joint design of such tools with local communities and cultural organizations, and establishing targeted standards regarding the quality and accuracy of cultural information to be made available. The strong demand identified among tourists coming from places far from the Mediterranean suggests that such tools could also serve as a differentiating factor in an increasingly crowded global market.

### 7.3 For yacht charter companies

Yacht charter companies are in a privileged position as intermediaries between sailing tourists and Mediterranean communities. They have access to their customers before departure – which is essentially a moment of high receptivity for cultural guidance. Charter companies could consider incorporating cultural briefings – to the extent possible they know their customers' destinations – and provide culturally focused recommendations that work with local communities to facilitate opportunities for deeper engagement. This data collection methodology used in the study – distributing questionnaires directly to customers on board – also demonstrates the value of charter companies as research partners in the ongoing study of sailing tourism behavior.

## 8. Limitations and Future Research Directions

There are limitations to this study that need to be acknowledged. First, the fact that the sampling is based on customers of a single yacht charter company may limit the general character of the findings to the broader population of sailing tourists. Future research could seek to replicate this study across multiple operators, destinations, and national contexts in order to derive more representative findings.

Second, the reliance of questionnaires on self-reported data introduces the potential for social desirability bias by respondents, particularly in relation to cultural sensitivity attitudes. Tourists may overestimate their commitment to cultural respect or underestimate uncertainty in order to present themselves favorably. Future studies cannot supplement the survey data with observational or behavioral monitoring methods to obtain more objective measures of cultural engagement, as it is important to maintain anonymity.

Third, while the collection window used adds temporal depth and reduces seasonal bias, it does not allow for a longitudinal analysis of individual tourist behavior over time. A study with the same tourists on multiple visits that could potentially yield valuable information on how cultural sensitivity evolves is practically impossible.

Future research could focus on the design and effectiveness of specific applications with cultural sensitivity, technical and operational convenience. Finally, corresponding comparative studies in different Mediterranean destinations — Greek, Italian, Croatian and Turkish port communities — would provide a richer understanding of the variation in needs.

## 9. Conclusion

This study aimed to examine the relationship between smart port city infrastructure, the use of digital tools and the cultural sensitivity and engagement of international sailing tourists in Greece. Through a survey of three tourist periods among 203 respondents, a set of empirical and practical findings emerged.

The greatest contribution of this study is the identification of a gap regarding tourists' cultural readiness for cultural respect and their actual uncertainty when confronted with local cultural norms. This gap stems from the absence of culturally integrated digital services and affects the growing segments of the sailing tourism market – namely visitors visiting a Mediterranean destination for the first time.

The conceptual framework of the study provides a blueprint for filling this gap. The demand for culturally aware smart port applications (88.1% positive responses, increasing to 91.2% among tourists with high uncertainty) demonstrates that the market for such tools is real and potentially urgent.

In conclusion, smart port applications in the Mediterranean have the opportunity to move closer to cultural intelligence. Integrating cultural awareness into the digital infrastructure of port tourism, thus improving the visitor experience, while simultaneously strengthening tourist relations with the community, thus contributing to a more humane and sustainable model of sailing tourism for the coming decades.